# Understanding the Near-Field Photoacoustic Spatiotemporal Profile from Nanostructures

Hanwei Wang[1,2], Yun-Sheng Chen[1,3,4,5,*], Yang Zhao[1,2,4,6*]

[1] Department of Electrical and Computer Engineering, University of Illinois Urbana-Champaign, Urbana, IL, USA

[2] Micro and Nanotechnology Laboratory, University of Illinois Urbana-Champaign, Urbana, IL, USA

[3] Beckman Institute for Advanced Science and Technology, University of Illinois Urbana-Champaign, Urbana, IL, USA

[4] Department of Bioengineering, University of Illinois Urbana-Champaign, Urbana, IL, USA

[5] Carle Illinois College of Medicine, University of Illinois Urbana-Champaign, Urbana, IL, USA

[6] Carl R. Woese Institute for Genomic Biology, University of Illinois Urbana-Champaign, Urbana, IL, USA

*Correspondence authors: yunsheng@illinois.edu, yzhaoui@illinois.edu

**ABSTRACT**

Understanding the mechanism of photoacoustic generation at the nano-scale is key to developing more efficient photoacoustic devices and agents. Unlike the far-field photoacoustic effect that has been well employed in imaging, the near-field profile leads to a complex wave-tissue interaction but is under-studied. Here we show that the spatiotemporal profile of the near-field photoacoustic waves can be shaped by laser pulses, anisotropy, and the spatial arrangement of nanoparticle(s). Using a gold nanorod as an example, we discovered that the near-field photoacoustic amplitude in the short axis is ~75% stronger than the long axis, and the anisotropic spatial distribution converges to an isotropic spherical wave at ~50 nm away from the nanorod's surface. We further extend the model to asymmetric gold nanostructures by arranging isotropic nanospheres anisotropically with broken symmetry to achieve a precisely controlled near-field photoacoustic "focus" largely within an acoustic wavelength.

**INTRODUCTION**

The photoacoustic effect, since discovered by Alexander Bell in 1880[1], has wide applications in spectroscopy[2-4], medical imaging[5-10], and recently neuroscience[11]. The photoacoustic effect generates acoustic waves through light absorption in an optically absorbing sample. Plasmonic nanostructures, including gold fabricated nanostructures[12] and synthetic nanoparticles[13-15], are popular photoacoustic materials[16], because they are chemically inert, have low toxicity, and, critically, their optical properties can be conveniently controlled by their geometries[17]. The plasmonic resonances of some anisotropic gold nanoparticles are located in the near-infrared (NIR) spectral region, also known as the tissue window, where light can penetrate deeper into tissue because of the minimum endogenous tissue absorption and scattering[18]. Due to the excellent optical properties and biocompatibility, these NIR gold nanoparticles play a critical role in developing theranostic agents[8, 19, 20], photothermal therapy[21-23], neuron stimulation[11, 24, 25], and photoacoustic molecular imaging technologies[16, 18, 26].

Driven by the great interest in gold nanoparticles in photoacoustic applications[11, 18, 24, 27-34], the fundamental mechanism of photoacoustic generation from various gold nanoparticles has been investigated theoretically and experimentally[18, 35-38]. The photoacoustic generation in gold nanoparticle solutions is distinct from the photoacoustic generation in pure light-absorbing solid or liquid. In gold nanoparticles, during the photoacoustic process, the photon energy converts into heat through thermal expansion and relaxation. Because of the high thermal conductivity of gold and the nanometer size, the generated heat leaks out of the nanoparticles during the light exposure and deposits most of the thermal energy in the surrounding media (usually the water or tissue), inducing photoacoustic waves from the surrounding media[36, 38, 39]. Distinctly, in pure bulk materials, heat is confined within the same medium because of the optical homogeneity. Compared to the bulk materials, the difference in confinement at the nanoscale leads to several unique photoacoustic properties. For example, different coating of the nanoparticles can affect the photoacoustic intensity[26, 36, 38, 40-42], albeit with the same optical absorption. Changing the size and shape of the nanoparticles affects the heat distribution, causing the different photoacoustic conversion efficiency, the efficiency of converting photon energy to acoustic pressure[18, 35, 43]. These properties have paved new design rules for producing nanoparticles with effective photoacoustic signal generation in the far-field.

Recently, there has been growing interest in photoacoustic devices[44-47], where the nanoscale stimulation using the photoacoustic effect differs dramatically from their applications in photoacoustic imaging. While the latter focuses on far-field photoacoustic waves, the former essentially utilizes the photoacoustic effect within the near-field of the nanoparticles and/or nanodevices. Others and our previous work have shown that gold nanoparticle aggregation causes a nonlinear effect[48, 49], which results in an enhancement of the photoacoustic fields and waves, suggesting that the nanoparticle near-field photoacoustic interactions may contribute to this nonlinear effect.

Several recent studies of the photoacoustic near-field have revealed that the near-field distribution around nanoparticles and microparticles differs from the far-field prediction[38] and may even have a spatial distribution depending on the shape of the particle[50]. However, the near-field spatiotemporal profiles remain largely understudied for anisotropic nanoparticles and nanostructures within the spatial region much smaller than the wavelength of the associated acoustic pulse. This knowledge gap might be attributed to the fact that many existing studies recorded the nanoparticle-generated photoacoustic signals with ultrasound transducers, where the near-field photoacoustic spatial profile was overlooked. However, the near-field photoacoustic profile is of paramount importance in the case of cellular stimulation and photoacoustic-induced drug release.

In this paper, we focus on analyzing the near-field photoacoustic profile of plasmonic nanostructures and start with a gold nanorod, a representative anisotropic plasmonic nanostructure[51]. Our results show that the photoacoustic spatial profile is anisotropic within a spatial domain that is much smaller the wavelength of the acoustic wave, yielding 74.6% higher signal intensity along the transverse direction of the nanorod than in the longitudinal direction on the nanoparticle surface. This near-field spatial profile depends on the sizes and geometries of the nanostructure and the laser pulse durations. Further, we show that the near-field photoacoustic profile can be manipulated by the anisotropic spatial arrangement of isotropic gold nanostructures.

**RESULTS AND DISCUSSION**

We use the following equation, $p(\mathbf{r},t)$, to describe the photoacoustic spatiotemporal profile[52],

$$(\nabla^2 - \frac{1}{v_s^2}\frac{\partial}{\partial t^2})p(\mathbf{r},t) = -\frac{\beta}{\kappa v_s^2}\frac{\partial^2 T(\mathbf{r},t)}{\partial t^2}, \tag{1}$$

where $v_s$ is the acoustic speed; $\beta$ is the thermal expansion coefficient; $\kappa$ is the isothermal compressibility, $\kappa = \frac{C_p}{\rho v_s^2 C_v}$; $C_p$ is the heat capacity at constant pressure; and $C_v$ is the heat capacity at constant volume.

We use a semi-analytical approach to analyze the near-field photoacoustic profile. The detailed three-dimensional thermal profile governs the near-field photoacoustic generation; therefore, it requires numerical calculations. However, the comparison with far-field photoacoustic signals requires an extensive computational burden. To reduce the computational burden, we combine numerical calculations and analytical approaches in the following.

We first numerically calculate the time-dependent thermal distribution of a gold nanorod. We choose to start with a gold nanorod because it has been widely used in far-field photoacoustic research[18, 26, 36, 49], is the simplest anisotropic nanostructure, and serves as a convenient model system. With a gold nanorod of 160 nm in length, 30 nm in width (i.e., an aspect ratio of 5.33), and water as the media, the absorption peak of the longitudinal mode of the nanorod falls in the second near-infrared (NIR-II) window (Fig. 1**a**). We simulate the nanorod absorption spectrum with a finite element solver (COMSOL Multiphysics 5.5, RF Module, Frequency Domain). The time dependent intensity $I$ of the laser pulse is modeled with a Gaussian function, $I = I_0 \cdot \exp(-\frac{1}{2}\frac{(t-t_0)^2}{\sigma^2})$, where $I_0$ is the peak intensity, $I_0 = \frac{F}{\sqrt{2\pi}\sigma}$; $F$ is the fluence of the laser; $\sigma$ is the variance $\sigma = \frac{w}{2\sqrt{2\ln 2}}$; $w$ is the full-width-at-half-maximum (FWHM) of the laser pulse; and $t_0$ is the center time of the pulse.

To begin with, we assume the laser fluence and pulse width to be 1 mJ cm$^{-2}$ and 0.1 ns, respectively; the fluence is kept constant throughout the paper and the pulse width is considered as a variable in our later analysis. The laser pulse width associates with the thermal confinement of the nanorod. If the pulse width is much shorter than the thermal relaxation time of the nanorod, the photoacoustic process is thermally confined; otherwise, the photoacoustic process is not thermally confined, where the heat transfer to the immediate media of the nanorod competes with the photothermal heating process. The thermal relaxation time of the nanorod, $\tau_t$, defined as the time required to dissipate $1-e^{-1}$ of the thermal energy with an impulse laser input[53], is 0.248 ns (Supplementary Note 1). If one assumes the thermal confinement, the heat transfer between the nanorod and its media is negligible, the second derivative of temperature in Eq. (1) can be simplified as

$$\frac{\partial^2 T}{\partial t^2} = \frac{1}{C_{nr}m_{nr}}\frac{\partial H}{\partial t}, \tag{2}$$

where $C_{nr}$, $m_{nr}$, and $H$ are the heat capacity, mass, and the total absorption power of the nanorod. $H = \sigma_{abs} I$, where $\sigma_{abs}$ is the absorption cross-section of the nanorod. The time derivative of the laser's intensity is $\frac{\partial I}{\partial t} = -\frac{(t-t_0)}{\sigma^2} P_0 \cdot \exp(-\frac{1}{2}\frac{(t-t_0)^2}{\sigma^2})$.

The assumed laser pulse width of 0.1 ns is smaller but comparable to the thermal relaxation time, therefore, the time-dependent heat transfer cannot be ignored. As shown in Fig. 1, the simulation considering the heating and cooling processes results in a lower photoacoustic amplitude than the analytical calculation (Eq. 2) by assuming thermal confinement (Fig. 1**b**). Fig. 1**c** shows the time evolution of the photoacoustic wave generated by the nanorod, which propagates through space with a far-field profile similar to a spherical wave but a distinct anisotropic feature in the near-field.

The anisotropy of the near-field photoacoustic distribution is attributed to the nanorod's geometry and acoustic interference. Neglecting the acoustic scattering at the nanorod/water interface, we can calculate the photoacoustic field using Green's function in water $G(\mathbf{r},t,\mathbf{r'},t') = \frac{\delta(t-t'-\frac{|\mathbf{r}-\mathbf{r'}|}{v_s})}{4\pi|\mathbf{r}-\mathbf{r'}|}$. The photoacoustic pressure can be simplified as

$$p(\mathbf{r},t) = \frac{\beta}{4\pi\kappa v_s^2}\int d\mathbf{r'}\frac{1}{|\mathbf{r}-\mathbf{r'}|}\frac{\partial^2 T(\mathbf{r'},t')}{\partial t'^2}\bigg|_{t'=t-\frac{|\mathbf{r}-\mathbf{r'}|}{v_s}}. \tag{3}$$

When the photoacoustic source region (the nanorod and its immediate media) is comparably small to their distance to the measurement point (i.e., far-field), $|\mathbf{r'}| \ll |\mathbf{r}|$, the term $\frac{1}{|\mathbf{r}-\mathbf{r'}|} \approx \frac{1}{|\mathbf{r}|}$. On the other hand, when the laser pulse width is relatively long, the time-shift of the photoacoustic wave generated from

different parts of the source region is negligible, $\frac{|\mathbf{r}-\mathbf{r'}|}{v_s} \ll t$, thus, $t' \approx t$. With these two approximations, the photoacoustic distribution can be further simplified into a spherical wave:

$$p_{sph}(\mathbf{r},t) = \frac{\beta}{4\pi\kappa v_s^2} \frac{1}{|\mathbf{r}|} \int d\mathbf{r'} \frac{\partial^2 T(\mathbf{r'},t)}{\partial t'^2}. \tag{4}$$

However, the spherical wave assumption fails to hold in the near-field. And when the pulse width is relatively short, the time-shift of the photoacoustic wave generated from different parts of the source region becomes significant; the positive acoustic pressure generated by one part may destructively interfere with a negative acoustic pressure caused by another. Both effects will lead to differences in the photoacoustic field distribution compared to the spherical wave described by Eq. (4).

As seen in Fig. 1**b**, the generated photoacoustic signal oscillates in time. To further analyze the spatial distribution of the photoacoustic near-field, we define the photoacoustic amplitude as the difference between the maximum and the minimum of the photoacoustic pressure over time, $P(\mathbf{r}) = \max(p(\mathbf{r},t),t) - \min(p(\mathbf{r},t),t)$. Note that the temporal profile of the photoacoustic pressure is asymmetrical under thermal non-confinement, so $\max(p(\mathbf{r},t),t) - \min(p(\mathbf{r},t),t) \neq 2\max(p(\mathbf{r},t),t)$. Fig. 2**a** shows the 2D photoacoustic amplitude distribution, $P(\mathbf{r})$, calculated with Eq. (3) using the simulated temperature profile. For comparison, Fig. 2**b** shows the photoacoustic amplitude distribution of a spherical wave $P_{sph}(\mathbf{r})$ given by Eq. (4).

The anisotropic spatial distribution in Fig. 2**a** motivates us to analyze the photoacoustic amplitude profiles along the transverse (x, or T) and longitudinal (y, or L) directions of the nanorod. With 0.1 ns laser pulse excitation, the pulse propagation length ($\Lambda$) defined as the FWHM pulse width multiplied by the acoustic speed $\Lambda = w v_s$ is 148 nm in water. This length scale is comparable to the length scale of the nanorod (160 nm); therefore, we anticipate discrepancies between the simulated photoacoustic amplitudes and the spherical wave estimated photoacoustic amplitude. Fig. 2c compares the simulated photoacoustic amplitudes along transverse and longitudinal directions with the semi-analytical solutions given by Eq. (3) and Eq. (4). It is seen that indeed the simulated photoacoustic amplitude deviates strongly from the far-field spherical wave estimation (Eq. (4)) when the distance to the nanorod surface is less than 50 nm, approximately 1/3 of the pulse propagation length $\Lambda$. This distance can be considered as the near-field regime where the far-field approximation breaks down, suggested by the convergence of all solutions in Fig. 2c to the spherical wave estimation beyond this distance. Fig. 2c also demonstrates that the simulated results and the semi-analytical Green's function method match well even within the near-field. The Green's function method estimates slightly larger photoacoustic amplitude than the simulated value because the former neglects the acoustic energy loss at the boundary of the nanorod. Nevertheless, the influence of this effect is minor, which has a difference smaller than 5% for 0.1 ns laser pulses and is expected to be even smaller for longer pulses, suggesting that the Green's function method can serve as a good approximation in the near-field regime.

To elaborate on the anisotropic near-field property of the nanorod, we define a unitless correction coefficient, $R_p$, which is the ratio of the photoacoustic near-field amplitude to the photoacoustic amplitude estimated from the spherical wave:

$$R_p(\mathbf{r}) = \frac{P(\mathbf{r})}{P_{sph}(\mathbf{r})} . \qquad (5)$$

Fig. 2d shows the 2D distribution of $R_p$ given by the ratio of Fig. 2a and Fig. 2b, highlighting the spatial anisotropy of the photoacoustic amplitude along the transverse and longitudinal directions. Fig. 2e analyzes how $R_p$ varies as a function of the distance away from the nanorod's surface along both directions, quantifying the differences between the near-field photoacoustic amplitude $P(\mathbf{r})$ and the amplitude estimated with spherical waves $P_{sph}(\mathbf{r})$ and thus, the anisotropy. On the nanorod surface, $P(\mathbf{r})$ is 20.0% higher than $P_{sph}(\mathbf{r})$ along the longitudinal direction but 58.5% lower along the transverse direction. It is worth noting that the absolute value of photoacoustic amplitude at the transverse location remains higher than at the longitudinal location because $|\mathbf{r}|$ is smaller at the width than the length of the nanorod and $\mathbf{r}$'s origin is the nanorod center. As a result, the photoacoustic amplitude at the transverse direction is 74.6% higher than the longitudinal direction on the nanorod surface, this is because there are more thermal sources contributing to the transverse direction than the longitudinal direction, which should not be confused with the electromagnetic field distribution shown in the inset of Fig. 1**a**.

The near-field photoacoustic amplitude can be further controlled by laser pulse widths due to heat transfer and acoustic interference. To fairly compare across different pulses widths, we use a normalized thermal energy $Q''w^2$. $Q''$ is the second derivative of the nanorod's thermal energy, where $Q$ is calculated by $Q = \rho C_{nr} \int_{nr} T dV$ and $w$ is the laser pulse width. $Q''w^2$ is proportional to the normalized photoacoustic pressure (see more details in Supplementary Note 2). Fig. 3**a** shows the comparisons of normalized thermal energy across different pulses widths; the N-shaped analytical solution is calculated as the time derivative of the absorbed power of the nanorod according to Eq. (2) under the thermal confinement condition. Therefore, the analytical solution shows the highest value. The normalized thermal energy is the higher with shorter pulse durations, which matches our expectations that the photoacoustic amplitude increases (Fig. S3**c**) as the laser pulse duration decreases because less thermal energy is leaked out to the surrounding medium, and it is approaching the analytical solution condition.

We conducted frequency analysis of the photoacoustic pulses to investigate how the laser pulse width changes the frequency responses of the photoacoustic signals. It is known that the photoacoustic pulses generated from nanoparticles contain a broad band of frequencies. The photoacoustic central frequency is tunable from several MHz to hundreds of GHz for laser pulses ranging from 0.01 ns to 100 ns

(Supplementary Note 3). While the high frequency components cannot propagation to the far-field, within the near-field regime, the duration of the laser pulse is a convenient approach to tune the frequency responses.

We discovered the different laser pulse widths also affect the near-field photoacoustic decay rate over distance. Note that here the decay rate is the initial photoacoustic amplitude distribution near a nanorod caused by thermal distribution, not the decay caused by the water attenuation of the propagating photoacoustic wave. According to equations (4) and (5), the distribution of the photoacoustic amplitude can be written as:

$$P(\mathbf{r}) = \frac{R_p(\mathbf{r})A_0}{|\mathbf{r}|}, \tag{6}$$

where $A_0$ is the source intensity defined as $A_0 = \max(A(t),t) - \min(A(t),t)$, and

$$A(t) = \frac{\beta}{4\pi\kappa v_s^2} \int d\mathbf{r}' \frac{\partial^2 T(\mathbf{r}',t)}{\partial t'^2}.$$

Since the photoacoustic amplitude will converge to the spherical wave (Eq. (4)) at a distance much larger than 50 nm, $R_p$ will approach 1 in the far-field for longer pulses ($w > 1$ ns). Based on Eq. (6), the near-field photoacoustic amplitude can be written as $P_{nf} = R_p(\mathbf{r_t})\frac{|\mathbf{r}|}{|\mathbf{r_t}|}P_{ff}$, where $P_{ff}$ and $P_{nf}$ are the far-field and near-field photoacoustic amplitude, $\mathbf{r_t}$ is the displacement with respect to the nanorod's center, and $R_p$ is evaluated in the near-field at the point of interest. $R_p$ reflects how fast the photoacoustic near-field decays. A higher $R_p$ with a certain far-field amplitude represents a higher near-field photoacoustic amplitude at the point of interest, in other words, a more rapidly decayed near-field in distance. By comparing $R_p$ in Fig. 3c and its spatial distribution in Fig. 3b, one can see that $R_p$ reaches its peak value with the pulse width between 0.1 ns and 1 ns.

When the pulse width is relatively short and the pulse propagation length $\Lambda$ is smaller than the nanorod's size, the positive and negative peaks of the photoacoustic wave generated by different parts of the source region destructively interfere with each other in the near-field, which will affect the near-field photoacoustic amplitude and change the photoacoustic pulse shape in the time domain (Fig. S2**a**, Supplementary Note 2). On the other hand, when the pulse width is longer than the thermal relaxation time, the heat transfer becomes significant, and the photoacoustic amplitude decreases as the pulse width increases. As both effects affect the near-field photoacoustic amplitude, our estimated $R_p$ using the Green's function method confirms a peak $R_p$ at the pulse width near the thermal relaxation time-scale and decreases toward longer and shorter pulse durations (Fig. 3**c**). Scrutinizing Fig. 3**c** suggests that to reach over 70% of the peak $R_p$, the laser should have a pulse propagation length longer than 0.46 of the nanorod's length, i.e.,

$$v_s \cdot w > 0.46 \cdot l, \tag{7}$$

and a pulse width shorter than 42.13 of the thermal relaxation time, i.e.,

$$w < 42.13 \cdot \tau_t. \tag{8}$$

To compare the spatial decay rate more intuitively, we fix the far-field photoacoustic amplitude at 1 μm and calculate the near-field with different pulse widths (Fig. 3**d**). Fig. 3**e** shows the photoacoustic near-field decays with a faster rate over distance for pulse widths of 0.1 ns and 1 ns than the other pulse widths, corroborating the optimal range of laser pulse widths for fast decayed photoacoustic near-field in Fig. 3**c**.

To optimize photoacoustic signals, many existing literatures have discussed the influence of physical dimensions and coating on the far-field photoacoustic signals[18, 26, 35-37]. Similarly, the physical dimensions and coating also affect the near-field photoacoustic amplitude. Both the photoacoustic equation (Eq. (1)) and the heat transfer equation (Supplementary Eq. (S5)) can be scaled up or down of its dependent variables without changing the correction coefficient, $R_p$. Eq. (7) and Eq. (8) remain valid regardless of the nanorod's size (Supplementary Note 4). We define a scaling factor, $a$, as the ratio of the nanorod's length to 160 nm (i.e., our initial nanorod length). When the nanorod is rescaled by a factor $a$, equivalently, the spatial variable is rescaled by $1/a$, and the temporal domain is rescaled by $a$. In this way, the heat transfer equation remains the same for different sizes of the nanorod so that one can optimize the nanorod dimensions while maintaining the nanorod's aspect ratio and thus, the optical resonance. During the photoacoustic generation, optical absorption efficiency and heat transfer will be affected by the nanorod dimension. Considering both effects, there is an optimal nanorod dimension for generating photoacoustic near-field amplitude at a given laser pulse width (Fig. S3, Supplementary Note 4).

In general, a smaller nanorod yields a higher absorption power density (Fig. S3**b**) but also a faster heat transfer, and consequently, a stronger cooling effect from the medium (Supplementary Note 4). Considering both effects, there is an optimal nanorod dimension for generating photoacoustic near-field amplitude at a given laser pulse width (Fig. 4). We observed that the optimal nanorod size is dependent on the laser pulse width and is approximately linear to the logarithm of the pulse width. Through linear fitting of the optimal scaling factor $a$, with the logarithm of the laser pulse width $w$ (Fig. S3**d**), the optimal scaling factor is approximately

$$a_{opt} = c_1 + c_2 \cdot \log(w/1ns), \qquad (9)$$

where the two coefficients are $c_1 = 0.0591$ and $c_2 = 0.402$. For example, for laser pulse widths of 0.1 ns, 1ns, and 10 ns, the optimal nanorod length is around 50 nm, 65 nm, and 80 nm (Fig. S3**c**), respectively. The photoacoustic amplitude is enhanced by 64.91%, 56.56%, and 48.36% accordingly compared to the nanorod with a length of 160 nm. Note that this is the near-field photoacoustic amplitude generated by a single nanorod with the assumption of a fixed laser fluence, distinct from that generated from an ensemble of nanoparticles in solution.

The near-field photoacoustic signal can be further enhanced with a coating, such as lipid[38, 42, 54] because the coating layer changes interfacial thermal resistance and the nanorod becomes more thermally confined at the same physical length. However, a thicker coating also increases the distances between the nanorod and the coating surface, which decreases the photoacoustic amplitude on the coating surface; therefore, there is an optimal coating thickness as well (Supplementary Note 5).

The overall optimal nanorod design and laser pulse width is shown in Fig. 4 to maximize the near-field photoacoustic amplitude by considering both the nanorod size and lipid coating thickness. Specifically, the shaded region between the black and red curves offers the rapidly decayed photoacoustic field, where $R_p$ is over 70% of its peak value. The relationship between the laser pulse width and the nanorod's size for the optimal photoacoustic amplitude is the blue curve (given by Eq. (9)). The scaling factor should be between 0.2 and 0.47, corresponding to a nanorod's length of 32 nm to 75.2 nm with a resonance in NIR II, and the laser pulse width should be between 0.01 ns and 4.6 ns according to these dimensions. The optimal thickness of a lipid coating should be between 8.5 nm and 25.6 nm, given by Eq. (S10) shown as the cyan curve in Fig. 4.

This abovementioned optimization methodology is not limited to gold nanorod but can be applied to near-field photoacoustic amplitude generated by general nanostructures. For example, the anisotropic near-field distribution can also be formed by isotropic gold nanospheres in an anisotropic arrangement. We discovered that a similar anisotropic photoacoustic near-field distribution as Fig. 1 can be formed using a linear array of three isotropic gold nanospheres (Figure S5, Supplementary Note 6). The field distribution and the decay rate highly depend on the arrangement of the nanostructure. The near-field anisotropy can be further manipulated by breaking the symmetry of the arrangement. As shown in Fig. 5**a**, when the top and bottom nanospheres are displaced by 30 nm toward the +x direction, the near-field photoacoustic pressure on the right side of the nanospheres becomes stronger than the left (Fig. 5**b**). The stronger photoacoustic amplitude at a certain location can be further enhanced by introducing more asymmetry in the nanostructure (Fig. 5**c** and 5**d**) and requires optimization.

In summary, we develop a systematic analysis of the near-field photoacoustic profile in this paper, starting from a representative anisotropic gold nanoparticle. The laser pulse width is a tunable parameter to control the decay rate of the photoacoustic near-field. One can consider the near-field photoacoustic spatial distribution as the spatial 'resolution' of the nanoparticle, where such a near-field photoacoustic signal extends beyond the nanoparticle region and interacts with the nanoparticle's immediate medium. Our results show that the near-field photoacoustic amplitude at a specific location can be optimized. The destructive acoustic interference reduces the near-field photoacoustic amplitude for short laser pulses, and thermal non-confinement reduces the photoacoustic amplitude for long laser pulses; therefore, for specific dimensions of the nanoparticle, there is an optimal laser pulse duration. Our study provides a general design rule for choosing the laser pulses, optimizing the sizes, and designing the anisotropy of a nanoparticle or a nanostructure to manipulate and achieve a designated near-field photoacoustic distribution for enhancing localized interactions.

## DATA AVAILABILITY

The datasets generated during and/or analyzed during the current study are available from the corresponding authors on reasonable request.

## ACKNOWLEDGEMENTS

The authors acknowledge funding supports from Jump ARCHES endowment through the Health Care Engineering Systems Center, Dynamic Research Enterprise for Multidisciplinary Engineering Sciences (DREMES) at Zhejiang University and the University of Illinois Urbana-Champaign, and NIGMS

# FIGURES

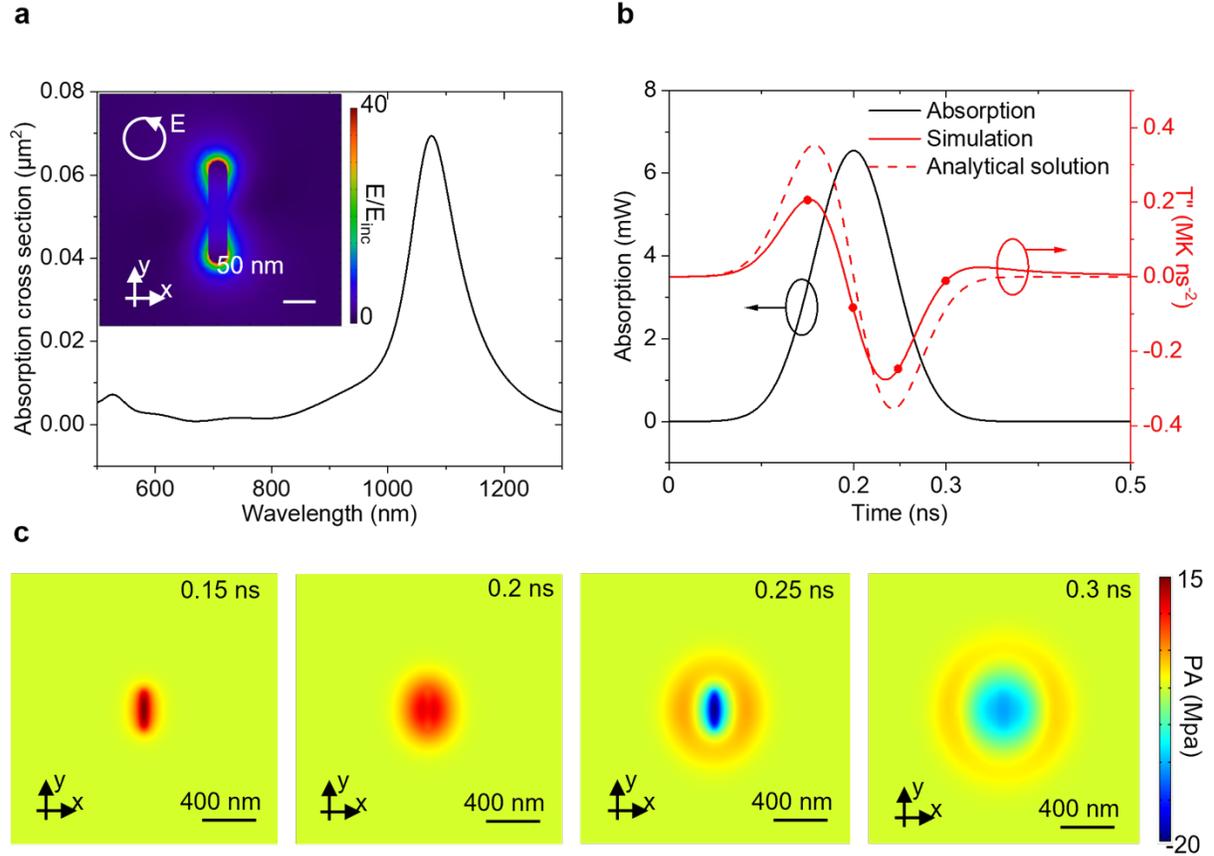

**Fig. 1| Calculated anisotropic photoacoustic near-field spatial distribution generated by a gold nanorod. a,** Optical absorption spectrum. The nanorod has a length of 160 nm and a width of 30 nm. The laser has a fluence of 1 mJ cm$^{-2}$ and a wavelength of 1100 nm. The laser is circularly polarized to excite both the longitudinal and transverse modes of the nanorod. Note that the particular handedness (left or right) is not critical regarding the nanorod's absorption. The inset shows the electric field distribution on resonance (at 1100 nm). The absorption spectrum and the electric field distribution are simulated with a finite element solver (COMSOL Multiphysics 5.5, RF Module, Frequency domain). **b,** Absorption and second derivative of temperature as a function of time. The temperature profile is simulated using a finite element solver (COMSOL Heat Transfer Module). **c,** Numerical simulation of the photoacoustic (PA) pressure wave using the partial differential Eq. (1), simulated with a finite element solver (COMSOL, PDE Module). The corresponding time frames are marked as the red dots in **b**.

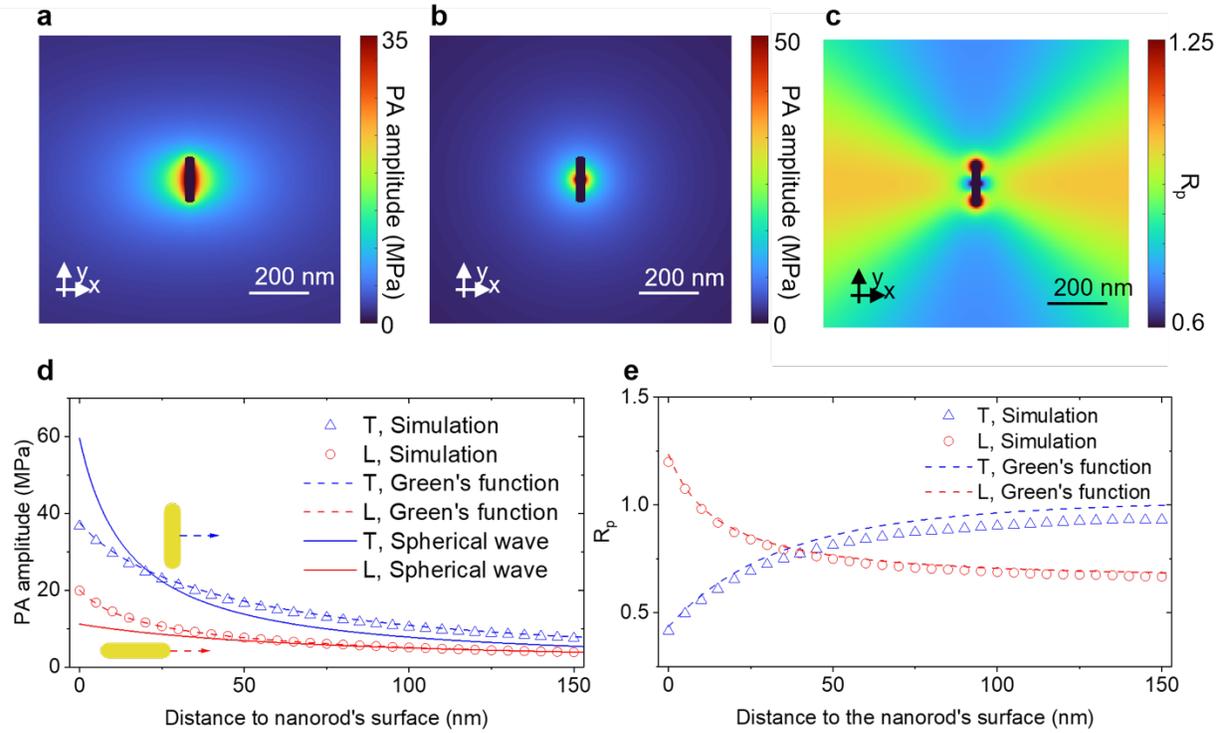

**Fig. 2| Influence of nanoparticle orientation on the near-field photoacoustic amplitude. a,** The near-field photoacoustic (PA) amplitude of a nanorod calculated by the Green's function method. The field inside the nanorod is set as zero. **b,** The photoacoustic amplitude of a spherical wave given by Eq. (4). The source of the spherical wave locates at the center of the nanorod. **c,** Photoacoustic amplitude calculated using Multiphysics numerical simulation. the Green's function method (Eq. (3)), and spherical wave approximation (Eq. (4)) along the transverse (T) and longitudinal (L) directions of the nanorod. The distance starts from the surface of the nanorod, as illustrated in the insets. **d,** 2D map of the correction coefficient $R_p$, which is the ratio between the photoacoustic amplitude in **a** to the spherical wave in **b**. **e,** $R_p$ along the transverse (T) and longitudinal (L) directions, calculated using the simulation and the Green's function method, showing good agreement in the near-field region.

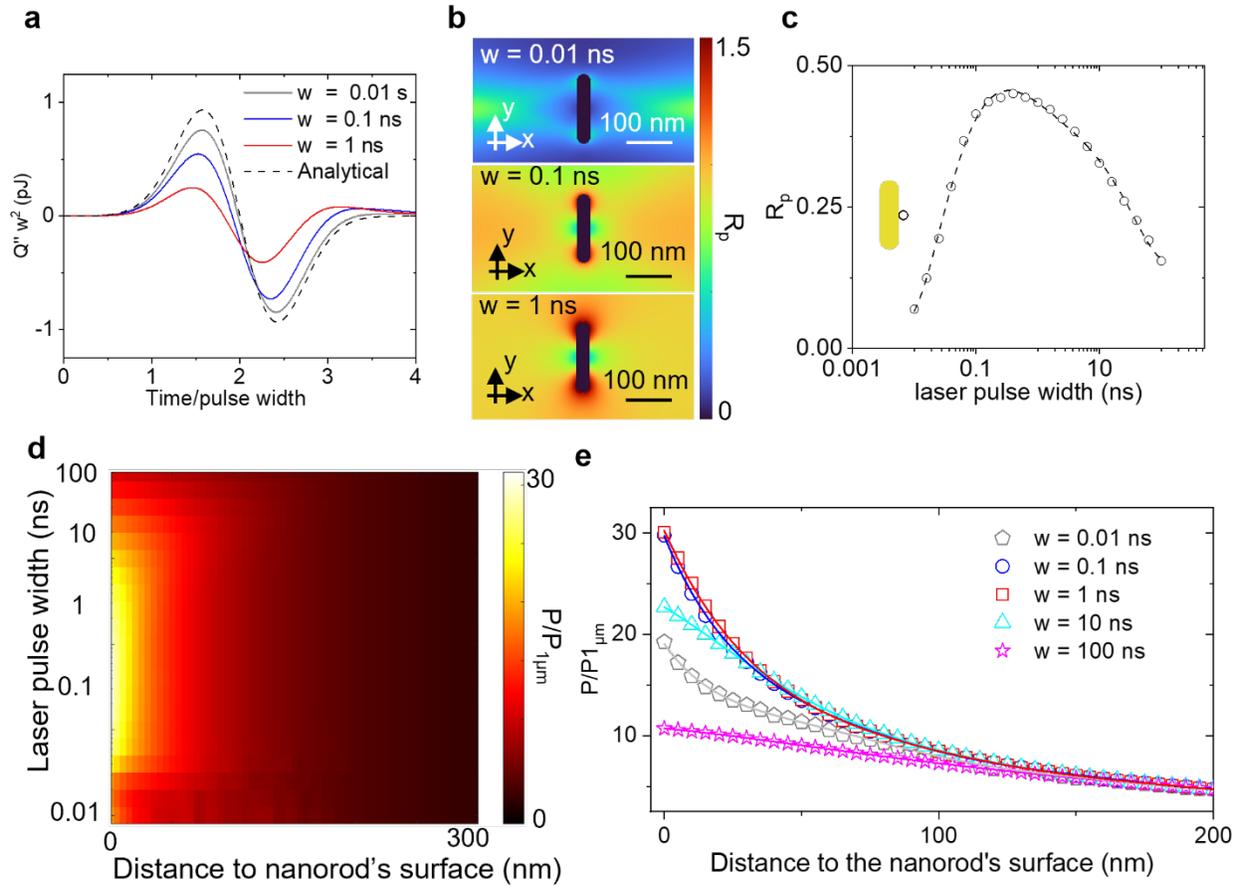

**Fig. 3| Tuning the near-field photoacoustic amplitude and decay rate with laser pulse widths. a,** The second time derivative of the nanorod's thermal energy $Q$ versus time with different pulse widths. The thermal energy is normalized by multiplying the square of the pulse width (Supplementary Note 2), and the time is normalized by dividing the pulse width. **b,** 2D map of $R_P$ with different pulse widths. **c,** $R_P$ on the surface of the nanorod (with the position marked in the inset) as a function of the pulse width, calculated using the Green's function method. For pulse widths longer than 1 ns, the time shift due to the size of the nanorod can be neglected. **d,** The photoacoustic pressure (P) normalized by photoacoustic pressure at 1 $\mu m$ (P$_{1\mu m}$) along the transverse direction as a function of the pulse width and the distance from the nanorod surface. **e,** Normalized photoacoustic pressure versus distance with different pulse widths showing the spatial decay rate.

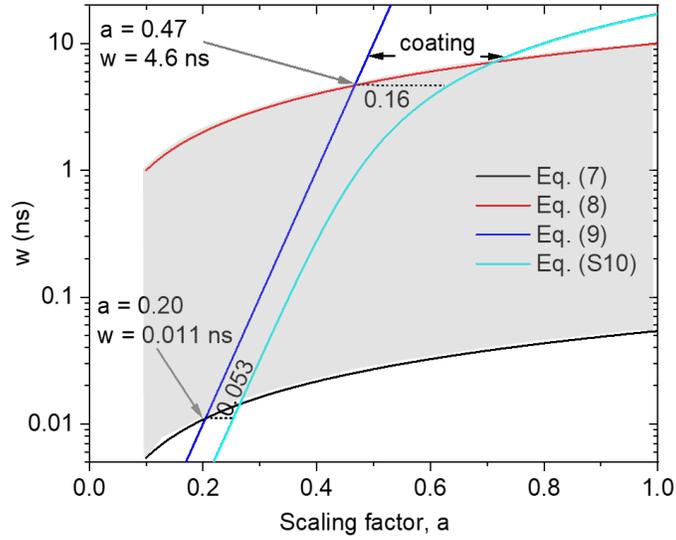

**Fig. 4| Design criteria of the laser pulse widths, nanorod materials and dimensions for optimizing near-field photoacoustic amplitude.** The black, red, blue and cyan curves are given by conditions in Eq. (7), Eq. (8), Eq. (9), and Eq. (S10), respectively. The shaded region between the black and red curves provides the highest decay rate of the photoacoustic field, and the blue curve provides the highest photoacoustic intensity. The blue curve between the two arrows marks the optimal condition considering both the decay rate and the photoacoustic intensity. The distance along x direction between the blue and the cyan curves shows the optimal coating thickness of a lipid shell. The shaded region illustrates the more rapidly decayed photoacoustic near-field, yielding $R_p$ over 70% of its peak value. $w$ denotes the laser pulse width; $a$ denotes the scaling factor of the nanorod.

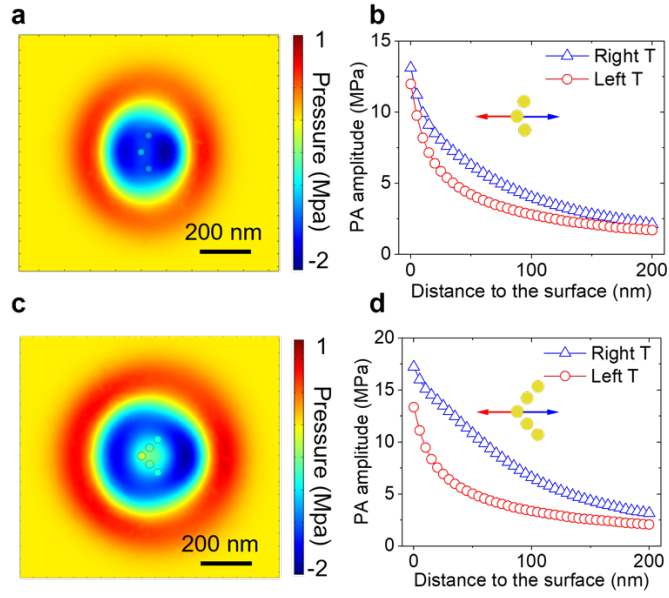

**Fig. 5| Manipulating the near-field photoacoustic distribution using anisotropic nanostructures. a,** Simulated photoacoustic (PA) amplitude at 0.35 ns generated by three nanospheres. The top and bottom spheres are displaced by 30 nm compared to the case in Fig. S3. The nanospheres have a radius of 15 nm and a separation of 65 nm. The laser pulse width is 0.1 ns. We use the same absorption power density as the nanorod at 1100 nm in Fig. 1a. **b,** Simulated PA amplitude along the transverse (T) direction. Blue curve shows the PA amplitude to the right of the nanostructure; the red curve shows the PA amplitude to the left of the nanostructure. **c,** Simulated PA amplitude at 0.35 ns generated by five nanospheres. The nanospheres have 25 nm separations in the y direction and 30 nm separation in the x direction, showing a strong anisotropy in the near-field PA spatial distribution. **d,** Simulated PA amplitude along the left and right transverse (T) directions for the five nanosphere case.

Supplementary information for **Understanding the Near-Field Photoacoustic Spatiotemporal Profile from Nanostructures**

Hanwei Wang[1,2], Yun-Sheng Chen[1,3,4,5,*], Yang Zhao[1,2,4,6*]

**Supplementary Note 1. Thermal relaxation time**

The thermal relaxation time is defined as the time required to cool to $e^{-1}$ (36.8%) of its initial thermal energy. We simulate the decay of the nanorod's temperature starting from a uniform temperature distribution, being equivalent to the decay following a laser impulse, which is assumed in photoacoustic imaging[1]. As shown in Figure S1, the nanorod reaches 36.8% of its initial temperature at 0.248 ns.

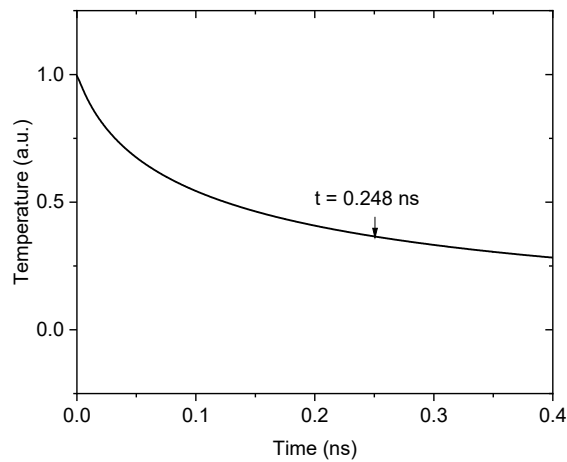

**Fig. S1| Cooling of the gold nanorod.** The nanorod starts with a uniformed temperature distribution and cools to 36.8% of its initial temperature at 0.248 ns.

**Supplementary Note 2. Normalization of nanorod's thermal energy under different laser's pulse width**

With a Gaussian pulse with a fixed fluence $F$, the optical absorption of the nanorod is

$$H(t) = \frac{F\sigma_{abs}}{\sqrt{2\pi}\sigma} \exp(-\frac{1}{2}\frac{(t-t_0)^2}{\sigma^2}), \tag{S1}$$

where the variance $\sigma$ is related to the full width at half maximum (FWHM) of the pulse $w$ as

$\sigma = \frac{w}{2\sqrt{2\ln 2}} = \frac{w}{2.3548}$. The time derivative of the optical absorption is

$$H'(t) = -2.3548^3 \frac{F\sigma_{abs}(t-t_0)}{\sqrt{2\pi}w^3} \exp(-\frac{1}{2}\frac{(t-t_0)^2}{\sigma^2}). \tag{S2}$$

Using $t/w$ as the variable, the parameter $H'(t)w^2$,

$$H'(t)w^2 = -2.3548^3 \frac{F\sigma_{abs}(t/w - t_0/w)}{\sqrt{2\pi}} \exp(-\frac{1}{2}2.3548^2(t/w - t_0/w)^2), \tag{S3}$$

is unrelated to the pulse width $w$. Therefore, we choose $H'w^2$ and $Q''w^2$ as normalized thermal energy of the nanorod. $Q''$ is the second derivative of the nanorod's thermal energy $Q$, calculated by $Q = \rho C_{nr} \int_{nr} TdV$ and $w$ is the laser pulse width. As the second derivative of the temperature, determined by the optical absorption, serves as source of the photoacoustic wave (Eq. (1) in the main text), we choose $Pw^2$ as the normalized acoustic pressure.

**Supplementary Note 3. Frequency spectra of the photoacoustic pulses**

The photoacoustic pulse in the time domain is shown in Fig. S2**a**, which approximately fits with the analytical solution (Eq. (2) in the main text) at a longer pulse width. The distortion at a shorter pulse width smaller than 0.1 ns is mostly attributed to the acoustic interference and a longer pulse width larger than 0.1 ns is mostly contributed by thermal non-confinement. The Fourier transform of the temporal profile of the photoacoustic pulse under thermal confinement (Eq. (2) in the main text) is

$$|F(p(t))| \sim \sigma f \exp(-\frac{1}{2}\sigma^2 f^2), \tag{S4}$$

where $f$ is frequency. The peak frequency of the spectrum (i.e., the central frequency of the pulse's spectrum) is $f_{peak} = \frac{1}{\sigma}$. Therefore, the peak frequency is reciprocal to the laser pulse width. A shorter laser pulse leads to a higher central frequency of the photoacoustic pulse. As the simulation shown in Fig. S2**b**, the simulated spectra of the photoacoustic pulses are similar to the analytical solutions, and the central frequency decreases with the laser pulse width. The differences are led by thermal non-confinement and acoustic interference. The effect of interference would be strong for particles comparably large to the acoustic pulse length, $wv_s$, such as the $w = 0.01$ ns case in Fig. S2. The photoacoustic pulse in the temporal domain becomes distorted from the N-shape pulse, and the spectrum becomes more complicated. For larger particles such as cells, this effect can be significant and the photoacoustic spectrum may contain more than one peak[2, 3].

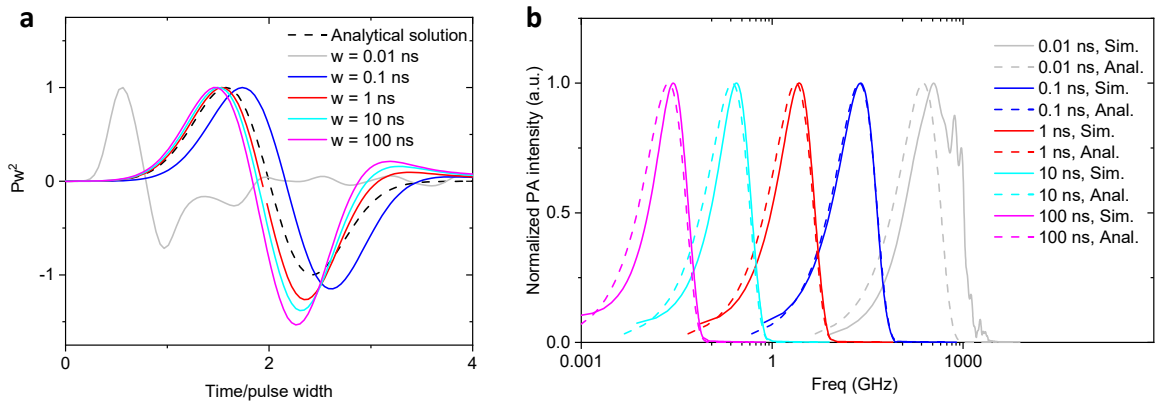

**Fig. S2| Tunability of photoacoustic signal in the frequency domain. a,** Normalized photoacoustic pressure (P$w^2$) in the time domain as a function of the time normalized to pulse widths (*w*). The temporal profile with laser pulse widths shorter or equal to 0.1 ns are calculated with Multiphysics simulations, and with widths longer than 0.1 ns are calculated with Eq. (1) in the main text using the simulated temperature profile. The analytical solution is given in Eq. (2) in the main text. **b,** Spectra of the photoacoustic pulse in **a** compared with the analytical solution. The analytical solutions are calculated by Eq. (S4). Sim. denotes numerical simulations; Anal. denotes analytical calculations.

**Supplementary Note 4. Rescaling of the thermal equation and photoacoustic equation**

The heat transfer equation of the nanorod is

$$\rho C \frac{dT(\mathbf{r},t)}{dt} = k\nabla^2 T(\mathbf{r},t) + H(\mathbf{r},t), \tag{S5}$$

where $\rho$, $C$, and $k$ are density, heat capacitance, and thermal conductivity. We use the parameters of gold in the nanorod region and water in the surrounding media, which can be taken as constants for linear thermal-elastics regime[4]. $H$ is the absorption power density. When the nanorod is rescaled by a factor $a$, equivalently, the spatial variable is rescaled by $1/a$, and the temporal domain is equivalent to be rescaled by $a$.

$$\rho C \frac{dT(\mathbf{r}_2,t_2)}{dt'} = k\nabla^2 T(\mathbf{r}_2,t_2) + H(\mathbf{r}_2,t_2), \tag{S6}$$

where $t_2 = at$, and $\mathbf{r}_2 = \frac{1}{a}\mathbf{r}$. For example, if the nanorod is reduced to half of its original size, $a = 0.5$. Instead of changing the nanorod's size, we define the new dependent variables as $t_2 = 0.5t$ and $\mathbf{r}_2 = 2\mathbf{r}$. In other words, the space is stretched to twice its original size, and the heat transfer process is sped up by two times, so the solution of the heat transfer equation remains the same.

The thermal relaxation time is therefore proportional to the scaling coefficient, $\tau_t = a\tau_{t_0}$, where $\tau_{t_0}$ is the thermal relaxation time of the nanorod with a length of 160 nm, $\tau_{t_0} = 0.248ns$. By choosing the normalized pulse width, $w_n$, as used in Fig. 3**c**, the thermal profile with respect to the normalized position, $\mathbf{r}/a$, remains constant regardless of the scaling factor.

Similarly, the photoacoustic equation (Eq. (1)) remains unchanged with the normalized position

$$(\nabla^2 - \frac{1}{v_s^2}\frac{\partial}{\partial t^2})p(\mathbf{r}/a,t) = -\frac{\beta}{\kappa v_s^2}\frac{\partial^2 T(\mathbf{r}/a,t)}{\partial t^2} \ . \tag{S7}$$

The photoacoustic pressure measured at the surface of the nanorod at the fixed normalized time $at$ remains the same per unit absorption power density. On the other hand, according to Eq. (6) in the main text, the far-field photoacoustic signal increases with the size of the nanorod (at a fixed concentration), which agrees with the literature[5].

Nanorods with different sizes yield various absorption efficiency, $Q_{abs} = \frac{\sigma_{abs}}{\sigma}$, where $\sigma$ is the physical cross-section, and $\sigma_{abs}$ is the absorption cross-section of the nanorod. The absorption efficiency varies marginally with different scaling factors $a$ (Fig. S3**a**). The absorption power density per unit volume is $W = \frac{\sigma_{abs}I}{v_{np}}$, and can be simplified as

$$W = \frac{Q_{abs}}{a}W_0 \ , \tag{S8}$$

where $W_0$ is the absorption power density with a nanorod's length of 160 nm.

The optimal nanorod size is dependent on the laser pulse width and is approximately linear to the logarithm of the pulse width. For example, for laser pulse widths of 0.1 ns, 1ns, and 10 ns, the optimal nanorod length is around 50 nm, 65 nm, and 80 nm (Fig. S3**c**), respectively. Due to the increased thermal energy loss to the media, the photoacoustic amplitude decreases with the pulse width. Note that the photoacoustic amplitude is different from $R_p$ discussed previously, which measures the ratio of the photoacoustic amplitude with the spherical wave amplitude. Both the photoacoustic amplitude and the $R_p$ decreases with the pulse width for long pulses, but with a different rate. Through linear fitting of the optimal scaling factor with the logarithm of the laser pulse width (Fig. S3**d**), the optimal scaling factor is

approximately

$$a_{opt} = c_1 + c_2 \cdot \log(w/1ns),  \quad (S9)$$

where the two coefficients are $c_1 = 0.0591$ and $c_2 = 0.402$. With the optimal dimension, the photoacoustic amplitude can be increased by 64.91%, 56.56%, and 48.36% respectively compared to the nanorod with a length of 160 nm.

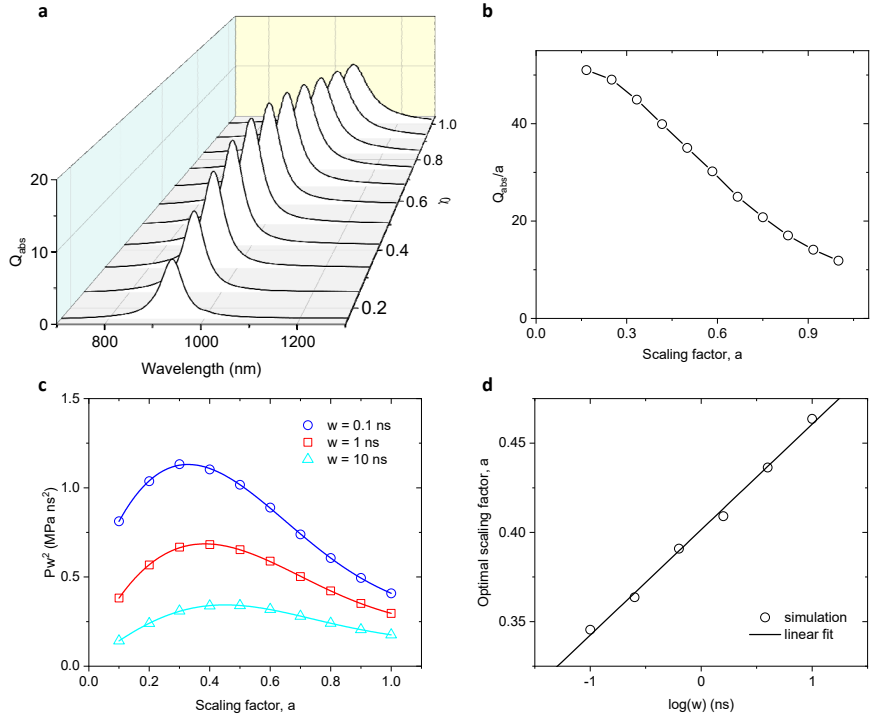

**Fig. S3| Optimize photoacoustic near-field amplitude by tuning nanorod dimensions. a,** The spectra of the absorption efficiency, $Q_{abs}$, with different scaling factors, $a$. **b,** $Q_{abs}/a$, being related to the absorption power density, versus the scaling factor, $a$. **c,** Photoacoustic amplitude at different scaling factors. The fluence with different pulse width is fixed at 1 mJ/cm$^2$. The photoacoustic amplitude $P$ is normalized by multiplying square of the pulse width, $w^2$ (Supplementary Note 2). The photoacoustic amplitude is measured at the spot marked in Fig. 3**c**. **d,** The optimal scaling factor versus logarithm of the laser pulse width.

**Supplementary Note 5. Optimizing photoacoustic near-field amplitude with coating**

The near-field photoacoustic signal can be further enhanced with a thermally isolative coating, such as lipid[6-8] because the nanorod becomes more thermally confined at the same physical length. The second time derivative of the temperature increases with the coating thickness (Fig. S4). However, the thicker coating also increases the distances between the nanorod and the coating surface, which reduces the photoacoustic amplitude on the coating surface. Due to these two effects, there is an optimal thickness of the shell layer, which is related to the laser pulse width (Fig. S4**b**). The optimal thickness is approximately linear to the pulse width (Fig. S4**c**). By fitting the simulation, we get the relationship as

$$a_{c,opt} = c_3 w + c_4, \qquad (S10)$$

where $a_{c,opt}$ is the ratio of the optimal coating thickness to the nanorod's length (as 160 nm), and the two coefficients are $c_3 = 0.025 ns^{-1}$ and $c_3 = 0.0487$.

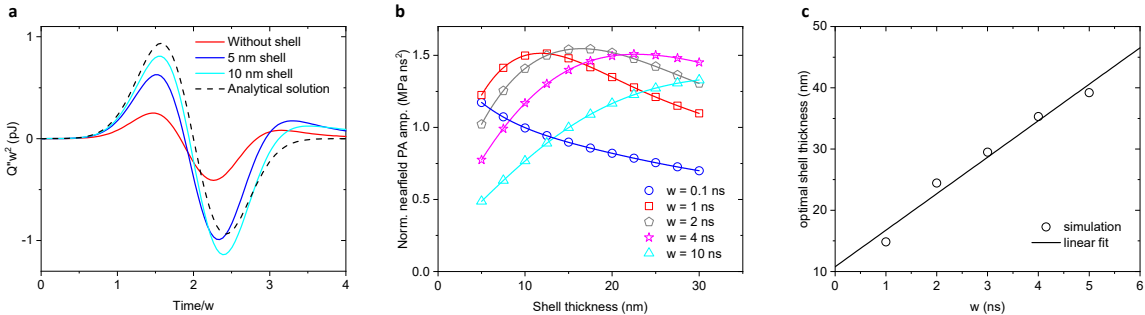

**Fig. S4| Tuning the near-field photoacoustic amplitude by lipid or dielectric coating. a,** The second time derivative of the nanorod's thermal power, Q, and the input power, H, with different coating thicknesses. The nanorod has a length of 160 nm, same as Fig. 2, 3, and 4. **b,** Photoacoustic amplitude versus shell thickness. The density, thermal conductivity, and heat capacity of lipid are 1.2 g cm³, 0.2 W m⁻¹ K⁻¹, 2348 J kg⁻¹ K⁻¹. The photoacoustic amplitude is normalized same as Fig. 6**c**. The photoacoustic amplitude is measured at the spot marked in Fig. 4**c**. The absorption of the nanorod is taken as a constant. **c,** The optimal shell thickness versus the laser pulse width. PA denotes photoacoustic.

**Supplementary Note 6. Anisotropic photoacoustic near-field distribution generated by isotropic nanoparticles**

Instead of anisotropic nanoparticles, isotropic nanoparticles arranged in anisotropic nanostructures can also create the anisotropic photoacoustic distribution. For example, Fig. S5**a** shows three nanospheres arranged linearly in the y direction. The second time derivative of the nanospheres' temperature is similar to that of the nanorod (Fig. 1**b,** main text). As a result, the photoacoustic amplitude created by the three nanospheres is also anisotropic (Fig. S5**b**), and the distribution is very similar to the one of the nanorod (Fig. 2**a,** main text). Similar to the nanorod's photoacoustic field, the photoacoustic field of the three nanospheres shows an anisotropic distribution in the near-field and converges to a spherical wave in the far-field (Fig. S5**c**).

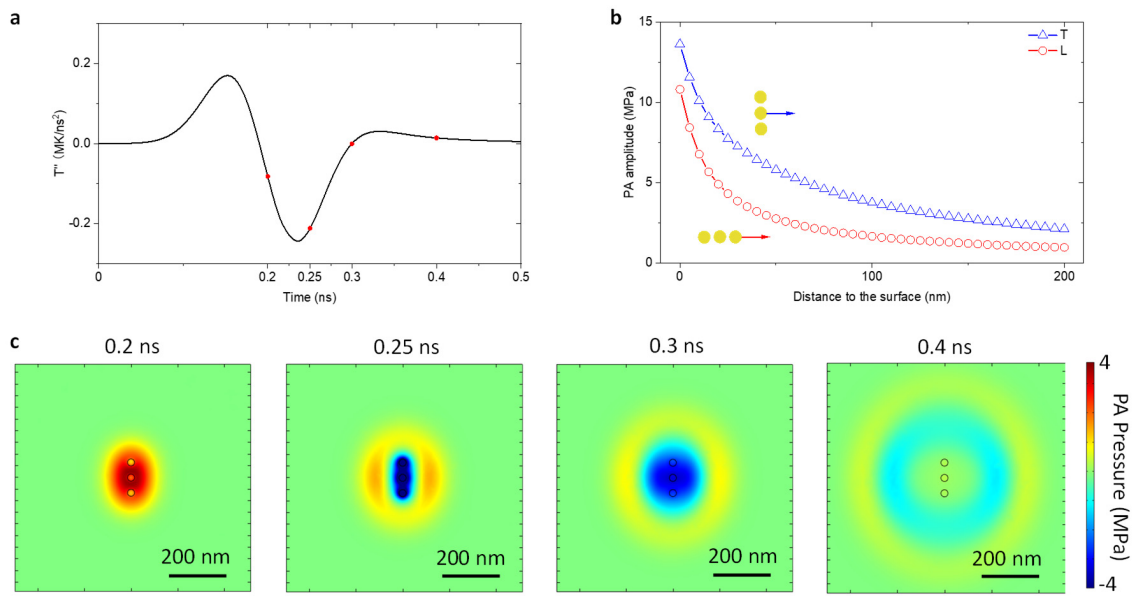

**Fig. S5| Anisotropic photoacoustic near-field distribution created by three isotropic nanospheres. a,** Simulated second time derivative of the nanosphere's temperature. The nanospheres have a radius of 15 nm and a separation of 65 nm. **b,** Distribution of the photoacoustic amplitude along the transverse and longitudinal directions. The transverse (T) and longitudinal (L) directions are illustrated in the inset. **c,** Simulated photoacoustic pressure at different time frames. The time frames are marked with the solid dots

in **a**. The laser pulse width is 0.1 ns. We use the absorption power density being the same with the nanorod at 1100 nm in Fig. 1**a** of the main text.